# Maria Reiche's Line to Archaeoastronomy

**Amelia Carolina Sparavigna**
Department of Applied Science and Technology
Politecnico di Torino, C.so Duca degli Abruzzi 24, Torino, Italy

Maria Reiche devoted her life to the study of the Nazca Lines, the most famous Peruvian geoglyphs. In fact, she was an archaeoastronomer that proposed for the Lines some interesting astronomical interpretations. We can appraise her approach using satellite imagery and a free planetarium software. A discussion of some geoglyphs is also proposed.

Maria Reiche was born on May 15th, 1903 in Dresden, Germany. Since her childhood, Maria was fond of natural sciences. In 1924, she started to study at the Dresden University of Technology, and after four years, she took the exams for teaching mathematics, physics, geography, and philosophy and pedagogy too. In the following years she had only temporary works, until she answered a job advertisement from the German consul in Cuzco, who was looking for a private tutor. In the February 1932 she travelled to Peru [1].

In 1940 Reiche became an assistant to Paul Kosok, a historian of the Long Island University, Brooklyn, New York. And then she started her field studies in the Nazca desert on the ancient "Nazca Lines", created from the Nazca culture probably before 1000 CE. These are the most famous geoglyphs of Peru, which are drawings on the ground representing animals, plants, lines, spirals and so on. Let us read Reiche's own words about the Nazca Lines [2].

*"The figures are of a whitish color on a brown surface, this brown surface is a thin covering of dark stone about 10 cm, which suffers the process of oxidation giving the entire region its particular brownish effect. Underneath the soil is still whitish, not brown, comprised of a mixture of rock that had been split into small fragments due to extreme temperatures, and clay, which ultimately was blown away by strong winds coming down from the Andes. ... There are extremely strong winds here, even sandstorms, but the sand never deposits over the drawings. On the contrary, the wind has a cleansing effect taking away all the loose material. This way the drawings were preserved for thousands of years. It is also one of the driest places on earth, drier then the Sahara. It rains only half an hour every two years! ... The figures, the drawings, are very superficial furrows never more than 30 cm in depth, and very shallow. For this reason the wind has obscured them by filling them with small dark pebbles from the surrounding surface like grain, making them difficult to detect from the air. To make them more accessible for viewing I cleaned them with a broom, one broom after another throughout the years. I went through so many brooms rumors circulated that I might be a witch!"*

Paul Kosok and Maria noticed that some of the lines were oriented to the sunrise on the winter solstice in the Southern Hemisphere. Therefore they began to map the lines for their relation to astronomical events. Around 1946, Maria Reiche began to map the figures and found different kinds of animals and birds. After Kosok left Nazca in 1948, she continued her researches in the area and proposed that the ancient Nazca people used them as an observatory for astronomical phenomena, publishing a book on her works entitled "The Mystery on the Desert" in 1949.

Maria devoted her life to the preservation of the Nazca desert and the lines. And concerns about the preservation of this archaeological site exist because the Pan American Highway is crossing it. Eventually, she convinced the government to restrict the access to the area. Moreover, she sponsored the building of a tower near the highway, so that an observer can have an overview of the lines without moving on and then damaging them.

Marie Reiche died on June 8, 1998.

Wikipedia [3], in the item on Maria Reiche, is telling that after the publication of her book on the lines, *"scholars concluded that the lines were not chiefly for astronomical purposes, but Reiche's and Kosok's work had brought scholarly attention to the great resource. It is widely believed that they were used as part of worship and religious ceremonies related to the calling of water from the gods."* In the item [4] on the Nazca Lines, it is told that *"Gerald Hawkins and Anthony Aveni, experts in archaeoastronomy, concluded in 1990 that there was insufficient evidence to support such an astronomical explanation."* Unfortunately, Wikipedia is giving as a reference to this phrase, a book written by Ian Cameron (Kingdom of the Sun God: A History of the Andes and Their People), not the original papers by Hawkins and Aveni.

As Giulio Magli wrote in Ref.[5], the Kosok and Reiche astronomical hypothesis was dismissed after a book published in 2002, written by H. Silverman and D.A. Proulx [6]. In this book the authors argued against the astronomical hypothesis, writing that "*an alignment between a celestial stars and a ground marking is statistically insignificant because countless stars are visible in the clear night sky at Nasca.*" Of course, we know very well that stars have different magnitudes, and then alignments can be statistical significant. Magli is also discussing the Hawkins and Aveni criticisms, after their original works. According to [5], Hawkins gained a conclusion studying only a few of the Nazca Lines.

In fact, Maria Reiche was an expert archaeoastronomer, because she applied her knowledge on mathematics and astronomy during on-site measurements; moreover her interpretation of the lines included their connection with the wet or rainy season. Of course, it is natural to imagine some worship on the place too. She is reporting her theory as follows [2].

"*In '46 I could see that the solstice lines existed in different places especially from centers, of which almost every one of them has one, or two, solstice lines. … More than sun directions there are moon directions, which is in agreement with the knowledge that the moon was observed before the sun…The geometric drawings are directed toward horizon points marking the rising and setting of the heavenly bodies and most likely served to mark the sowing and harvest time, and the distribution of food during the dry period of the year. The figures indicated the division of the year by way of constellations, with respect to their positions at night. The most important epoch of the year was, until now, December. This was the month the rivers would fill to the brim with muddy water that brought life to the fields. Now this has all stopped. There is an eternal drought here due to the contamination of air quality preventing the clouds from reaching the high mountains to fill the rivers. Years ago one could see the people making furrows in the fields to prepare for the arrival of the water. In ancient times they knew when to begin this labor … Here the Big Dipper (our Big Bear) announces the water. This constellation is only visible between December and March and is seen here upside down with the handle curved upward. It's possible that the Dipper was represented by one of the large drawings - the monkey. … and several straight lines point to the rising and setting of the largest star in the Dipper in the year 900 AD."*

Using the Google maps, we can see the Monkey and the parallel lines (see Figure 1). An angle of 27° with respect the North-South direction can be obtained (I suggest to assume an uncertainty of at least one degree). Using the planetarium software Stellarium (www.stellarium.org), setting the location at Nazca and the time on 900 CE, we see the star Alkaid, which is at the end of the tail of the Big Bear, setting more or less with the same azimuthal angle. Therefore what Maria is telling seems to be quite good.

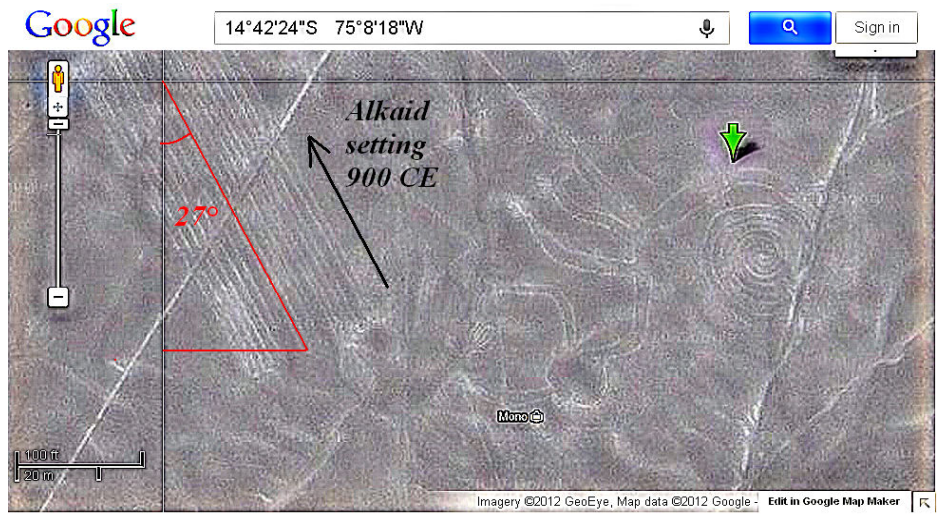

**Fig.1** Using the Google maps, we can see the Monkey and the parallel lines. An angle of 27° with respect the North-South direction can be obtained. Using the Stellarium (www.stellarium.org), setting the location at Nazca and time on 900 CE, we see the star Alkaid setting more or less on the same azimuthal angle

Maria Reiche is also mentioning the existence of triangles. Using the Google maps to survey the Nazca Lines, we immediately see the large quantities of lines. And the crossing of these lines creates triangles. Let us choose one of them, that shown in the Figure 2. We see that this image is geometrically interesting. One of the sides has a West-East direction approximately. The two other sides are quite visible. One of them is slightly widening, forming an angular sector.

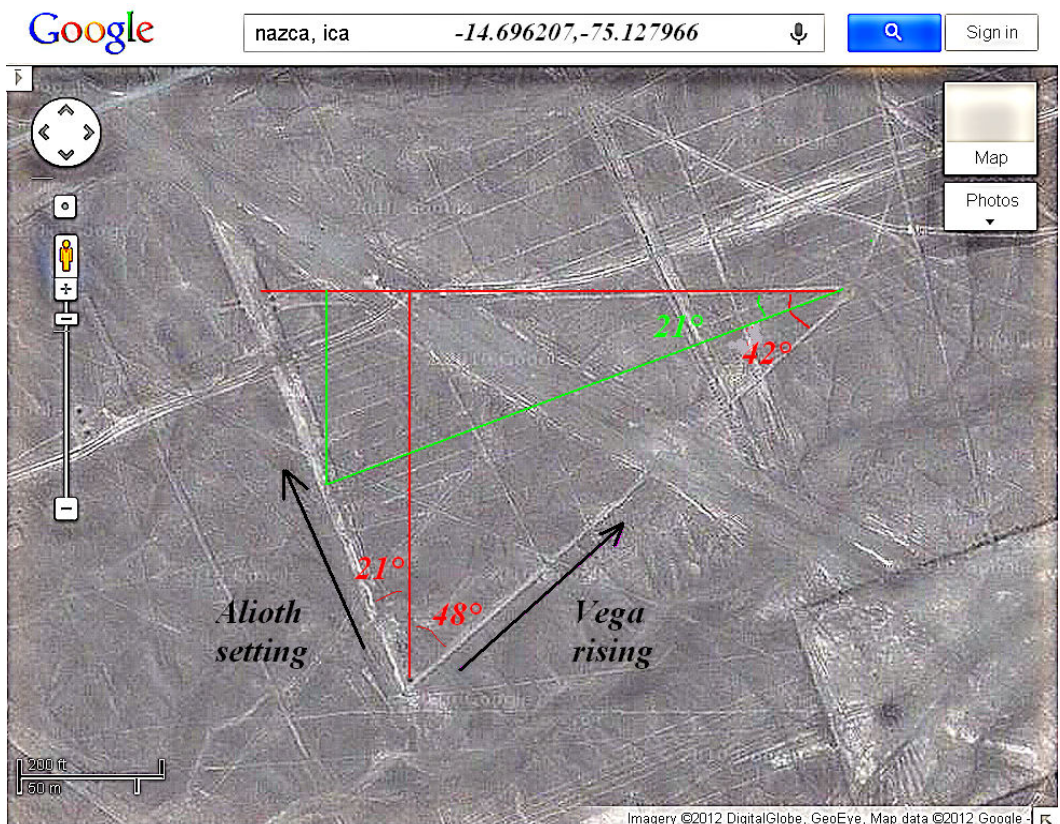

**Fig.2** A Nazca triangle. One of the sides has a West-East alignment approximately. The two other sides are quite visible. One of them is slightly widening, forming an angular sector. If we consider an uncertainty of two degrees, the directions of the two sides correspond to the rise of Vega and the setting of Alioth, on 1000 CE.

Let us consider an astronomical hypothesis and search for some stars rising or setting according to these sides, assuming an observer at the southern corner of the triangle (the satellite map shows there what seems to be a small mound). To this purpose we can use again the Stellarium. We run it setting a date of 1000 CE. The two directions of the sides seem to correspond to the rise of Vega and the setting of Alioth, in the Big Bear again. Assuming an uncertainty of the angles measured in the Google Maps of about two degrees and considering that the local visibility can be different due to some surrounding hills, it seems possible again an astronomical orientation.

According to Maria Reiche, besides the orientation of some lines with sun, moon and stars, the ancient people at Nazca had their constellations, and some of the geoglyphs were representing these constellations. And in fact, using Stellarium set on 900 CE towards North, we can see the Big Bear rotating about the celestial pole as Maria Reiche described. Moreover, the connection she made between astronomical phenomena and the wet season in Nazca is in agreement with all the proposed theories on the Nazca Lines. Therefore, it is not reasonable a total dismissal of a theory which proposes the lines possessing an orientation towards the rise and setting of celestial bodies and the position of celestial poles.

Let me discuss two other examples using the Google maps again.

Besides the lines, some of the geoglyphs representing animals are visible in the Google Maps. For instance, the Frigatebird (see Fig.3) is one of the huge geoglyphs which are properly appreciated when we observe them in the satellite images. This bird has a so long bill that is clearly representing a direction. As I have recently discussed [7], the bill was oriented towards the setting of Fomalhaut, between 800-1000 CE. Due to the precession of the Earth axis, the bill is no more pointing toward this star. The direction changed of 5°.

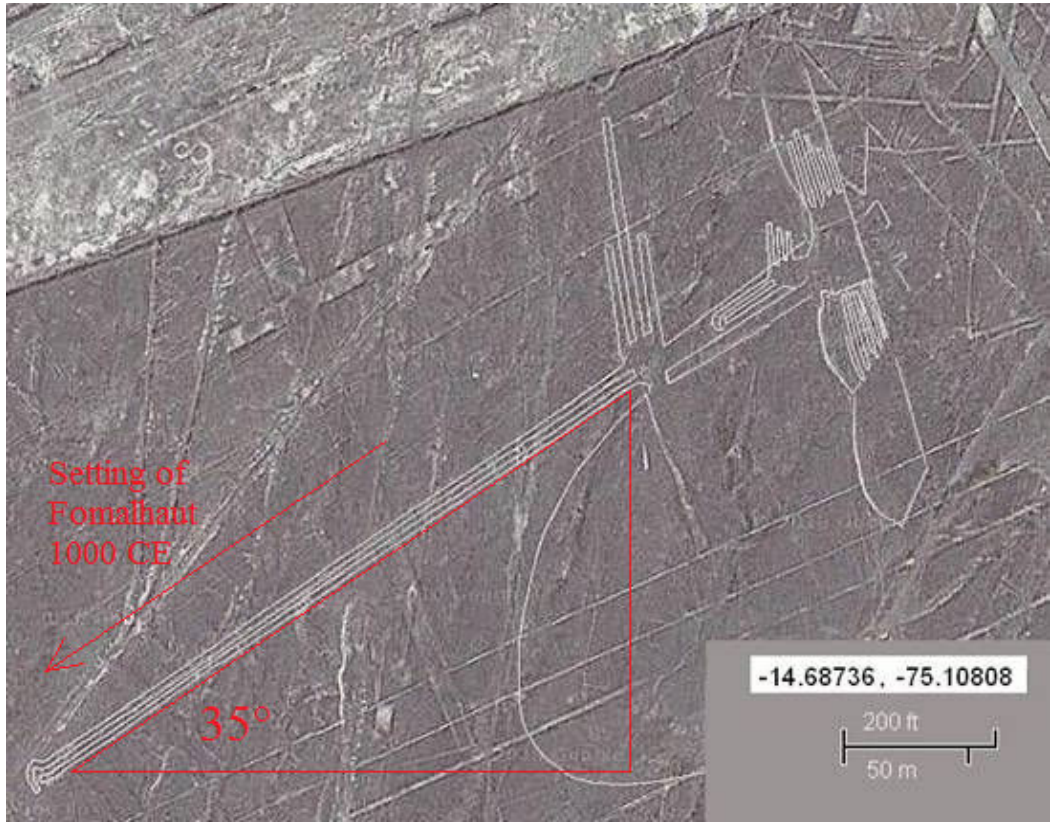

**Fig.3 The Nazca Frigatebird. His bill was pointed towards the setting of Fomalhaut in the Piscis Austrinus.**

My proposal [7] was that the Frigatebird had represented a bird snatching his food from the ocean surface, in the same manner as the geoglyph is snatching the star from the horizon. Note that Fomalhaut is the alpha star of the Piscis Austrinus. Then, some of the Nazca geoglyphs can be viewed as linked to constellations, a Maria Reiche suggested. However, these geoglyphs are not depicting the constellations in a strict sense, that is, they are not drawings that show the apparent relative positions of the stars. I prefer telling that the geoglyphs of animals can be some representations of the life on Earth projected in the sky. Therefore, the frigatebirds feeding themselves on the ocean are represented by the Frigatebird geoglyph that catches Fomalhaut at its setting after the run in the night sky, as if it were a fish in the ocean.

In the case of the Monkey (Fig.1), the animal has a huge spiral tail; this tail could be representing the rotation of the Big Bear about the celestial pole. Therefore the Monkey is representing the fact that the sky seems to rotate about the Earth axis.

Let me conclude with another example about this idea.

One of the most famous geoglyphs is that of the Hummingbird. As most hummingbirds, the geoglyph has a long and straight bill, a sharp bill adapted for drinking nectar from flowers. In the Figure 4, which is showing the Google Map of it, we do not see flowers near the Hummingbird: we see parallel long lines. The direction of these lines corresponds to the direction of the sunrise on the December solstice. This means that the Hummingbird is drinking the light of Sun. Let us compare this image with the words that the Peruvian writer and poet José María Arguedas wrote. He imagined "picaflores que llegan hasta el sol para beberle su fuego", that is hummingbirds flying toward the sun to drink his fire.

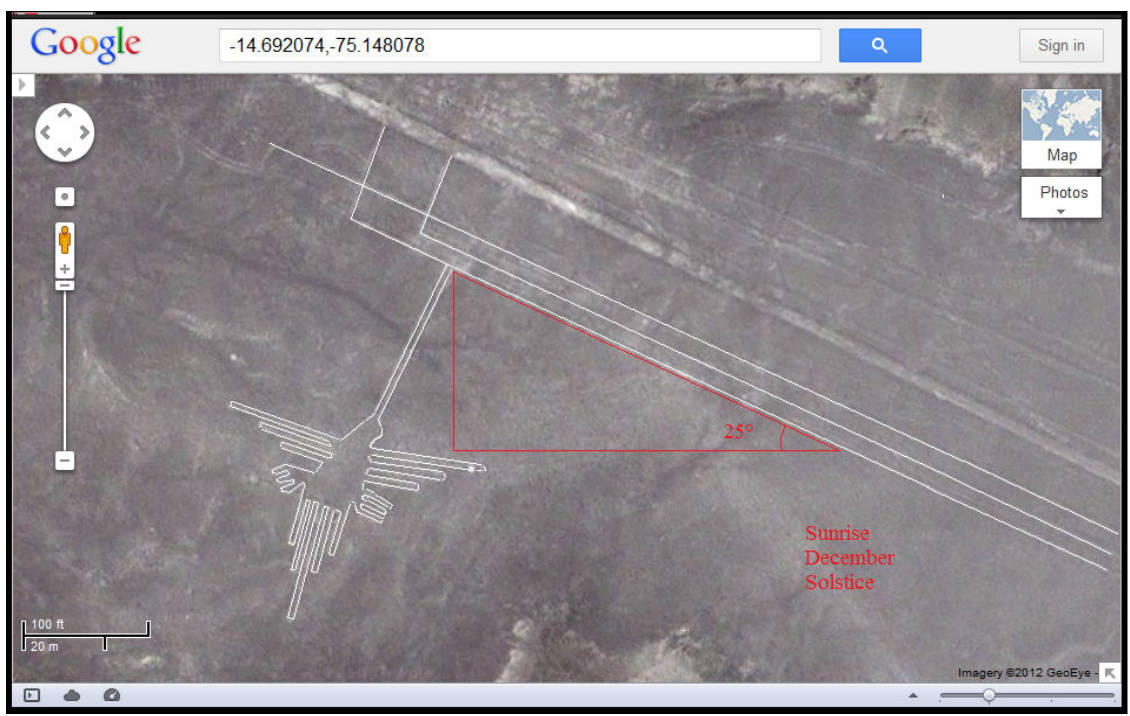

**Fig.4 The geoglyph of the Hummingbird has the bill touching a line having the direction of the sunrise on the December solstice. The Peruvian writer and poet José María Arguedas imagined "picaflores que llegan hasta el sol para beberle su fuego", that is, hummingbirds flying toward the sun to drink his fire.**

It would be interesting a comparison of the data recorded by Maria Reiche and the geoglyphs observed in satellite maps, in order to investigate other alignments of lines towards the rise and setting of stars. Some drawings are difficult to see in satellite imagery, but a suitable processing [8] could be enough to enhance them for this purpose. Moreover, the simulation of the sky locally observed and the effect of precession can be easily performed, using some planetarium software, freely available for all the researchers curious of archaeoastronomical facts.